\documentclass{ptephy}

\usepackage{amsmath} 
\usepackage{amsthm} 
\usepackage{cite} 
\usepackage{graphicx} 
\usepackage{algorithmic} 
\usepackage{subfig} 
\usepackage{url} 
\usepackage{latexsym}
\usepackage{natbib}
\usepackage{bm}
\bibliography{sample}

\begin{document}
\title{Fluctuations of the cosmic background radiation appearing in the $10$-dimensional 
cosmological model  }
\author{\name{K. Tomita}{\ast}}
\address{\affil{}{Yukawa Institute for Theoretical Physics, Kyoto University, Kyoto 606-8502, Japan}
\rm{\email{ketomita@ybb.ne.jp}}}

\begin{abstract}
We consider a cosmological model starting from 
(1) the $(1+3+6)$-dimensional space-times consisting of the outer space (the 
$3$-dimensional expanding section) and the inner space (the
$6$-dimensional section) and reaching (2) the Friedmann model after the decoupling of 
the outer space from the inner space, and derive fluctuations of the background radiation 
appearing in the above $10$-dimensional space-times. For this purpose  
we first derive the fluid-dynamical perturbations in the above $10$-dimensional space-times,
 corresponding to
two kinds of curvature perturbations (in the scalar mode) in the non-viscous case, and
next study the quantum fluctuations in the scalar and tensor modes, appearing at the
stage when the perturbations are within the horizon of the inflating outer space.
Lastly we derive the wave-number dependence of fluctuations (the power spectrum) in the
two modes, which formed at the above decoupling epoch and are observed in the Friedmann
stage. It is found that this can be consistent with the observed spectra of the cosmic microwave
background radiation.

\end{abstract}

\maketitle

\section{Introduction}

In order to derive the observed fluctuations of cosmic microwave background radiation,
we study the cosmological evolution of the $(1+3+6)$-dimensional space-times, in
which it is assumed that our universe was born as an isotropic and homogeneous
10-dimensional space-times and evolved to the state consisting of  the 3-dimensional
inflating outer space and the 6-dimensional collapsing inner space. Our 4-dimensional 
Friedmann universe appeared after decoupling of the outer space from the inner space. 
This scenario is supported by the present super-string theory  (Kim et al. \citep{kim1,kim2} 
in a matrix model). 

 In a previous paper\citep{tom}  we discussed the entropy production at the stage 
 when the above inflation and collapse coexist, and showed how viscous processes help 
 the increase of cosmological entropy. We also discussed the possibility that we satisfy, 
 at the same time,
 the condition that the entropy in the Guth level\citep{guth} is obtained and the 
 condition that the inner space decouples from the outer space.
 In the subsequent paper\citep{tom2} , 
 we studied the evolution of cosmological perturbations
  in the non-viscous case, solving the equations for geometrical perturbations. 
 
 In this paper we treat the fluidal perturbations corresponding to the geometrical
 perturbations, the quantum fluctuations in the scalar and tensor modes, and the
 consistency with the observations of cosmic microwave background radiation (CMB)
 in the non-viscous case.
 In Sect. 2, we first review the background model and the perturbed quantities.
 In Sect. 3, we derive the perturbed fluid-dynamical equations, corresponding to
 the geometrical perturbations in the 10-dimensional space-times, and solve them. 
 In Sect. 4, we consider the quantum fluctuations in the scalar and tensor modes at the 
 stage when they were within the horizon of the outer space with the inflationary expansion,
 and derive the initial conditions for their perturbations appearing after this stage.
 In Sect. 5, we derive the spectra of perturbations in these two modes, and compare
 them with the observed ones.  In Sect. 6, concluding remarks are given.    
 In Appendix A, we derive the higher-order terms in the two curvature perturbations with 
respect to small wave-numbers in the outer space.

\section{Review of the background model and the perturbed quantities}
\subsection{Our background model}
We consider a cosmological model starting from 
(1) the $(1+3+6)$-dimensional space-times consisting of the outer space (the 
$3$-dimensional expanding section) and the inner space (the
$6$-dimensional section), and reaching (2) the Friedmann model after the decoupling of 
the outer space from the inner space, as shown schematically in Fig. 1.

\begin{figure}[t]
\caption{\label{fig:bkr1} Scale factors of outer and inner spaces in the 10-dimensional
space-times and the Friedmann model. $t_{dec}$ and $t_A$ denote the decoupling epoch
(when the Friedmann model starts) and the singular epoch of the inner space, respectively.}
\includegraphics[width=8cm]{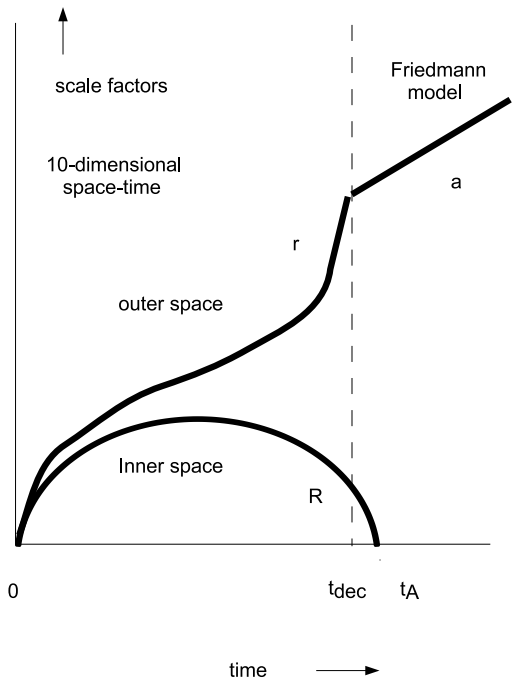}
\end{figure}

\subsubsection{Background 10-dimensional model before the decoupling}
The background 10-dimensional space-time is expressed in the form of a product of two
 homogeneous spaces $\rm{M_d}$ and $\rm{M_D}$ as
\begin{equation}
  \label{eq:a01}
ds^2 = -dt^2 + r^2(t)\ {}^d g_{ij} (x^k) \ dx^i dx^j + R^2(t)\ {}^Dg_{ab} (X^c) \ dX^a dX^b,
\end{equation}
where ${}^dg_{ij}$ and ${}^Dg_{ab}$ are the metrics of the outer space $\rm{M_d}$ and the 
inner space $\rm{M_D}$ with constant curvatures $K_r$ and $K_R$, respectively. Here the 
dimensions of $\rm{M_d}$ and $\rm{M_D}$ are $d = 3$ and $D = 6$. 
The inner space $\rm{M_D}$ expands initially and collapses after the maximum expansion with
$K_R = 1$, while the outer space  $\rm{M_d}$ continues to expand with $K_r = 0$ or $-1$. 
Then the background metric is 
\begin{equation}
  \label{eq:a02}
\begin{split}
g_{00} &= -1, \quad g_{01} = g_{0a} = g_{ia} = 0, \\
g_{ij} &= r^2\ {}^d g_{ij}, \quad g_{ab} = R^2\ {}^D g_{ab}, 
\end{split}
\end{equation}
and the Ricci tensor is
\begin{equation}
  \label{eq:a03}
\begin{split}
R^0_0 &= - \left(d \frac{\ddot{r}}{r} + D \frac{\ddot{R}}{R}\right), \\
R^i_j &= -\delta^i_j \ \left[\left(\frac{\dot{r}}{r}\right)^. + \frac{\dot{r}}{r} \left(d \frac{\dot{r}}{r} +
D \frac{\dot{R}}{R}\right) + (d-1) \frac{K_r}{r^2}\right], \\
R^a_b &= -\delta^a_b \ \left[\left(\frac{\dot{R}}{R}\right)^. + \frac{\dot{R}}{R} \left(d \frac{\dot{r}}{r} +
D \frac{\dot{R}}{R}\right) + (D-1) \frac{K_R}{R^2}\right],
\end{split}
\end{equation}
where $i, j = 1, ..., d, \ a, b = d+1, ..., d+D$, and an overdot denotes $d/dt$. 
At the singular stage when $R$ is near $0$, the curvature terms with $K_r/r^2$ and 
$K_R/R^2$ are negligible, compared with the main terms, and the curvatures can be treated 
 approximately as $K_r = K_R = 0$.
The background energy-momentum tensor is
\begin{equation}
  \label{eq:a04}
T^\mu_\nu = p \delta^\mu_\nu + (\rho + p) u^\mu u_\nu,
\end{equation}
where $u^\mu$ is the fluid velocity, $\rho$ the energy density, and $p$ the pressure. Here 
$\rho$ and $p$ are the common photon density and pressure in both spaces. 
The fluid is extremely hot and satisfies the equation of state $p = \rho/n$ of photon
gas, where $n = d + D = 9$. Einstein equations  are expressed as
\begin{equation}
  \label{eq:a05}
R^\mu_\nu = - 8\pi \bar{G} (T^\mu_\nu - \frac{1}{2} \delta^\mu_\nu T^\lambda_\lambda),
\end{equation}
where $\bar{G}$ is the $(1+d+D)$-dimensional gravitational constant. In the following, we 
set $8\pi \bar{G} = 1$. The background equation of motion for the matter is
\begin{equation}
  \label{eq:a06}
\frac{\dot{\rho}}{\rho+p} + d \frac{\dot{r}}{r} +D \frac{\dot{R}}{R} = 0.
\end{equation}

The Einstein equations for $r$ and $R$ were solved numerically in the previous paper\citep{tom}
and their behaviors were shown in Figs. 1 - 7 of [3].  
At the early stage, the expansion of the total universe is nearly isotropic (i.e. $r \propto R$). 
At the later stage, the inner space collapses after the maximum expansion, and at the final stage 
we have an approximate solution
\begin{equation}
  \label{eq:a07}
r = r_0 \ \tau^\eta,  \quad  R = R_0 \ \tau^\gamma  \quad (r_0, \ R_0 : const)
\end{equation}
with%
\begin{equation}
  \label{eq:a08}
\eta = \{1 - [D(n-1)/d]^{1/2} \}/n , \quad \gamma =  \{1 + [d(n-1)/D]^{1/2} \}/n,
\end{equation}
and $\tau = t_A - t$, where $t_A$ is the final time corresponding to $R = 0$. For $d = 3$ 
and $D= 6$, we have
\begin{equation}
  \label{eq:a09}
\gamma = - \eta = 1/3.
\end{equation}
For the solutions (\ref{eq:a07}), Eqs. (\ref{eq:a04}) and (\ref{eq:a05}) lead to $R^0_0 = 0$ and 
$T^0_0 - \frac{1}{2} T^\mu_\mu \propto \rho$, so that we have 
\begin{equation}
  \label{eq:a010}
\rho = 0
\end{equation}
at the final stage. The curvature tensor is singular, on the other hand, in the limit 
$\tau \rightarrow 0$, like that in the $4$-dimensional Kasner space-time.\citep{wald}

\begin{figure}[t]
\caption{\label{fig:bkr2} Ratios of physical sizes of perturbations with the wave-number 
$k$ to the Hubble length $1/H$. The ratio $r(\tau) H/k $ in the outer space of the
10-dimensional space-times ($\tau \equiv t - t_A$)  is shown as the solid line on the left-hand
 side, and the ratio $a(t_f) H/k $ in the Friedmann model is shown as the solid line
 on the right-hand side.}
\includegraphics[width=8cm]{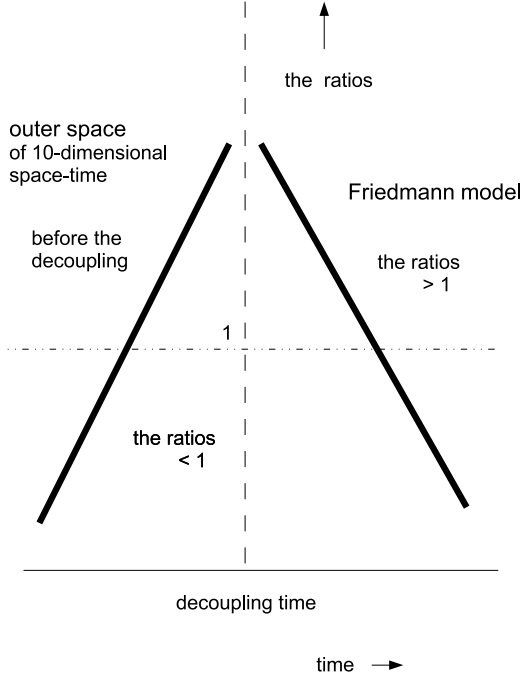}
\end{figure}

\subsubsection{Decoupling of the two spaces and the Friedmann stage}

As $\tau$ decreases, the inner space contracts and finally the size reaches the Planck
 length at the decoupling epoch. We discussed the decoupling condition at this
 quantum-gravitational epoch and the entropy 
 production to this epoch in the previous paper [3]. 
 
 At present we cannot analyze the process of decoupling accurately, because no
 quantum theory of gravitation has been established yet. However it is expected that
 the inner space is so homogeneous and quietly evolves without violent phenomena.
 This is because in both the inner and outer spaces the perturbations are assumed to be 
 caused by quantum fluctuations before the decoupling, grow gravitationally, and
 remain  very small and at the linear stage, before the decoupling. Note here that 
 gravitational instability in the outer and inner spaces was treated in the previous paper [5]. 
 Thus the inner space separates quietly from the outer space and disappears, while  
 in the outer space the Friedmann model appears after the decoupling  
 
 After this decoupling epoch it is assumed here that the outer space is separated from 
the inner space and described using the Friedmann model with the metric
\begin{equation}
  \label{eq:a010a}
ds^2 = -d{t_f}^2 + a^2 (t_f) [d\chi^2 + \sigma^2(\chi ) (d\theta^2 + \sin^2 \theta d\Omega^2)]
\end{equation}
with the cosmic time $t_f$ and $\sigma (\chi) = \sin \chi, \chi, \sinh \chi$ for the 
curvature $+, 0, -$ in the space, 
and the scale-factor $a (t_f)  \propto {t_f}^{1/2}$ at the radiation-dominated hot stage. 
At the decoupling epoch $t_{dec}$ and $(t_f)_{dec}$, the entropy is assumed to be conserved
in the 10-dimensional space-time and the Friedmann model.
The behavior of scale-factors is shown in Fig. 1.

\subsubsection{Horizon crossing}
The ratio of physical sizes of perturbations with the wave-length $1/k$ to the Hubble length
$1/H$ is 
\begin{equation}
  \label{eq:a010b}
r(\tau) H/k \propto \tau^{-4/3} = (t_A -t)^{-4/3}
\end{equation}
before the decoupling, where $r \propto \tau^{-1/3}$ and $H \propto \tau^{-1}$, and 
it increases with time.

After the decoupling epoch we have the  ratio at the Friedmann stage,
\begin{equation}
  \label{eq:a010c}
a(t_f) H/k  \propto {t_f}^{-1/2},
\end{equation}
which decreases with time. The two ratios are nearly equal at the decoupling epoch.

These ratios are shown schematically in Fig. 2. They can take the value $1$ in both sides 
before and after the decoupling 
epoch, that is, we can have the horizon crossing in both sides. Quantum fluctuations are 
created at the epoch of $r(\tau) H/k < 1$ in the outer space of the 10-dimensional 
space-time, and the fluctuations are observed as the fluctuations of the  background 
radiation at the epochs of $a(t_f) H/k < 1$ at the Friedmann stage.

\subsection{Perturbed quantities}
The simplest treatment of perturbations of geometrical and fluidal quantities is to expand 
them using harmonics, and to find the gauge-invariant quantities.  For the four-dimensional 
universe (in the Friedmann model) it was shown in Bardeen's theory on perturbations\citep{bar}. 
In the multi-dimensional universe  consisting of 
the outer and inner homogeneous spaces $\rm{M_d}$ and $\rm{M_D}$ with different geometrical
 structures, we can have no harmonics in the $(d + D)$-dimensional space. 
Abbott et al.\citep{abb}  considered separate expansions in $\rm{M_d}$ and $\rm{M_D}$
using the harmonics defined in the individual spaces,  classified the perturbations in $\rm{M_d}$ 
and $\rm{M_D}$ individually as scalar (S), vector (V), and tensor (T), and classified the 
6 types of perturbations in $\rm{M_d} + \rm{M_D}$ into three modes: the scalar mode 
(including SS), the vector mode (including SV, VS, VV), and the tensor mode (including ST, TS).
The left and right sides of signatures correspond to the types of perturbations in $\rm{M_d}$ and 
 $\rm{M_D}$, respectively. 
 
 In this paper only scalar and tensor modes are considered in the 10-dimensional space-times.
  So quantities in these modes are shown here. 
 
  \subsubsection{The scalar mode}
 The metric perturbations are expressed as
\begin{equation}
  \label{eq:a011}
\begin{split}
g_{00} &= - (1 + 2 A \ q^{(0)} Q^{(0)}),\\
g_{0i} &= - r b^{(0)} q_i^{(0)} Q^{(0)},\quad g_{0a} = - R B^{(0)} q^{(0)} Q_a^{(0)},\\
g_{ij} &= r^2 [(1+2h_L q^{(0)} Q^{(0)})\ {}^dg_{ij} + 2h_T^{(0)}q_{ij}^{(0)} Q^{(0)}],\\
g_{ab} &= R^2 [(1+2H_L q^{(0)} Q^{(0)})\ {}^Dg_{ab} + 2H_T^{(0)}q^{(0)} Q_{ab}^{(0)}],\\
g_{ia} &= 2rR G^{(0)} q_i^{(0)} Q_a^{(0)},
\end{split}
\end{equation}
where $q^{(0)}, q_i^{(0)}, q_{ij}^{(0)}$ and  $Q^{(0)}, Q_a^{(0)}, Q_{ab}^{(0)}$ are scalar harmonics in  $\rm{M_d}$ and $\rm{M_D}$, respectively, and $A, b^{(0)}, B^{(0)}, h_L, H_L, h_T^{(0)}, H_T^{(0)}, $
 and $G^{(0)}$ are functions of $t$.

 The perturbations of fluid velocities and the energy-momentum tensor are expressed as  
\begin{equation}
  \label{eq:a012}
u^0 = 1 - A q^{(0)} Q^{(0)},\quad
u^i = \frac{v^{(0)}}{r} q^{(0)i} Q^{(0)},\quad
u^a = \frac{V^{(0)}}{R}  q^{(0)} Q^{(0)a},
\end{equation}
and
\begin{equation}
  \label{eq:a013}
\begin{split}
T^0_0 &= - \rho ( 1+ \delta \ q^{(0)} Q^{(0)}),\\
T^0_i &= r (\rho + p)(v^{(0)} - b^{(0)}) q_i^{(0)} Q^{(0)},\quad
T^0_a = R  (\rho + p)(V^{(0)} - B^{(0)}) q^{(0)} Q_a^{(0)},\\
T^i_j &= p (1 +\pi_L q^{(0)} Q^{(0)}) \delta^i_j,\quad
T^a_b = p (1 +\Pi_L q^{(0)} Q^{(0)}) \delta^a_b,\\
T^i_a &= 0,
\end{split}
\end{equation}
where we consider a perfect fluid, so that the anisotropic pressure terms
vanish and we have
\begin{equation}
  \label{eq:a014}
\pi_L  = \Pi_L = \delta.
\end{equation}

For the metric perturbations in Eq. (\ref{eq:a011})  the following gauge-invariant quantities  
are defined:
\begin{equation}
  \label{eq:a015}
\begin{split}
\Phi_h &= h_L + \frac{h_T^{(0)}}{d} + \frac{r}{k_r^{(0)}} \frac{\dot{r}}{r} b^{(0)} - 
\frac{r^2}{k_r^{(0)2}}\frac{\dot{r}}{r} \dot{h}_T^{(0)},\\
\Phi_H &= h_L + \frac{H_T^{(0)}}{D} + \frac{R}{k_R^{(0)}} \frac{\dot{R}}{R} B^{(0)} - 
\frac{R^2}{k_R^{(0)2}}\frac{\dot{R}}{R}\dot{H}_T^{(0)},
\end{split}
\end{equation}
\begin{equation}
  \label{eq:a016}
\begin{split}
\Phi_A^{(r)} &= A + \frac{r}{k_r^{(0)}}\dot{b}^{(0)} + \frac{r}{k_r^{(0)}} \left(\frac{\dot{r}}{r} +
D\frac{\dot{R}}{R}\right) b^{(0)} \\
&- \frac{r^2}{k_r^{(0)2}}\left[\ddot{h}_T^{(0)} + \left(2\frac{\dot{r}}{r} +
D\frac{\dot{R}}{R}\right) \dot{h}_T^{(0)}\right] + D\left(H_L +  \frac{H_T^{(0)}}{D}\right),\\
\Phi_A^{(R)} &= A + \frac{R}{k_R^{(0)}} \dot{B}^{(0)} + \frac{R}{k_R^{(0)}} \left(d\frac{\dot{r}}{r} +
\frac{\dot{R}}{R}\right) B^{(0)} \\
&- \frac{R^2}{k_R^{(0)2}} \left[\ddot{H}_T^{(0)} + \left(d\frac{\dot{r}}{r} +
2\frac{\dot{R}}{R}\right) \dot{H}_T^{(0)}\right] + d\left(h_L +  \frac{h_T^{(0)}}{d}\right).
\end{split}
\end{equation}
The gauge-invariant quantities $\Phi_h$ and $\Phi_A^{(r)}$ in the outer space 
correspond to the gauge-invariant perturbations defined by Bardeen\citep{bar} in the
$(1+3)$-dimensional usual universes, and  $\Phi_H$ and $\Phi_A^{(R)}$ in the inner 
space are similar to the above quantities. $\Phi_h$ and  $\Phi_H$ represent the 
curvature perturbations in both spaces.

The gauge-invariant quantities for fluid velocity and energy density perturbations are given by
\begin{equation}
  \label{eq:a017}
v_s^{(0)} = v^{(0)} - \frac{r}{k_r^{(0)}} \dot{h}_T^{(0)}, \quad
V_s^{(0)} = V^{(0)} - \frac{R}{k_R^{(0)}} \dot{H}_T^{(0)}, 
\end{equation}
and
\begin{equation}
  \label{eq:a018}
\epsilon_m = \delta +\frac{n+1}{n}\left[d\frac{\dot{r}}{k_r^{(0)}} (v^{(0)}-b^{(0)})+
D\frac{\dot{R}}{k_R^{(0)}} (V^{(0)}-B^{(0)})\right] .
\end{equation}
It should be noticed that $v_s^{(0)}, V_s^{(0)}$ and $\epsilon_m$ do not vanish, though $\rho = 0$ 
at the final stage as in Eq. (\ref{eq:a010}). 
As a gauge-invariant quantity that has no counterpart in the usual universe, we have
\begin{equation}
  \label{eq:a019}
\Phi_G = G^{(0)} - \frac{1}{2} \frac{k_R^{(0)}}{k_r^{(0)}}\frac{r}{R} h_T^{(0)} - 
 \frac{1}{2} \frac{k_r^{(0)}}{k_R^{(0)}}\frac{R}{r} H_T^{(0)}.
\end{equation}
Moreover the auxiliary quantities ($\Phi_6$ and $\Phi_7$) and $\tilde{\Phi}_G$ are defined by
\begin{equation}
  \label{eq:ag2}
\frac{\dot{r}}{r} \frac{\dot{R}}{R} \Phi_6 \equiv \frac{\dot{R}}{R} \left(h_L +\frac{h_T^{(0)}}{d}\right) -
\frac{\dot{r}}{r} \left(H_L +\frac{H_T^{(0)}}{D}\right), 
\end{equation} 
\begin{equation}
  \label{eq:a4a}
  \Phi_7 \equiv (r/\dot{r}) \Phi_h -  (R/\dot{R}) \Phi_H - \Phi_6,
\end{equation}
and
\begin{equation}
  \label{eq:ag1}
\tilde{\Phi}_G \equiv \frac{rR}{k_r^{(0)} k_R^{(0)}} \Phi_G.
\end{equation}
\subsubsection{The tensor mode}
We have only metric perturbations given by
\begin{equation}
  \label{eq:a027}
\begin{split}
g_{00} &= - 1, \quad g_{0i} = g_{0a} = g_{ia} = 0, \\ 
g_{ij} &= r^2 ({}^d g_{ij} + 2h_T^{(2)}q_{ij}^{(2)} Q^{(0)}),\\
g_{ab} &= R^2  ({}^D g_{ab} + 2H_T^{(2)}q^{(0)} Q_{ab}^{(2)}),
\end{split}
\end{equation}
and have no fluidal perturbations, where have we neglected anisotropic stresses.
In this mode, $h_T^{(2)}$ and $H_T^{(2)}$ correspond to the TS and ST parts of curvature
 perturbations and they themselves are gauge-invariant.
 
\medskip  
More details about perturbations can be seen in the previous paper [5].

\section{Evolution of fluidal perturbations in the scalar mode }

In the previous paper [5], we derived the equations for geometrical perturbations 
$\Phi_h, \Phi_H, \tilde{\Phi}_G$ and $\Phi_6$ in the 10-dimensional space-times, 
and found their behavior by solving them.  In this section we derive
the equations for gauge-invariant variables representing fluidal perturbations 
$\epsilon_m, v_s^{(0)}$ and $V_s^{(0)}$ from the equations $\delta T^\nu_{\mu; \nu} = 0$ with  $\mu
 = 0, i$ and $a$, respectively, and derive their behaviors, where the suffices $\nu, i$ 
 and $a$ take the values $0 \sim d+D,  1 \sim d$ and 
$d+1 \sim d+D$, respectively, in the outer and inner spaces with dimensions $d$ and $D$,
respectively, where $d = 3$ and $D = 6$.  
In the following, $v_s^{(0)}$ and $V_s^{(0)}$ are expressed as $v_s$ and $V_s$ for 
simplicity.

First we obtain the following equation for $\dot{\epsilon}_m$ from $\delta T^A_{0; A} = 0$ 
\begin{equation}
  \label{eq:a1}
 \begin{split} 
\frac{n}{n+1} &[\dot{\epsilon}_m - \frac{1}{n} (d\frac{\dot{r}}{r}+D\frac{\dot{R}}{R}) \epsilon_m]\\
&= -d\{\dot{\Phi}_h +[D\frac{\dot{R}}{R}+ (d-2)\frac{\dot{r}}{r}]\Phi_h \} +[-\frac{k_r^{(0)}}{r}
 +\frac{d}{k_r^{(0)}}(\ddot{r}-\frac{\dot{r}^2}{r}) ] v_s \\
& -D\{\dot{\Phi}_H +[d\frac{\dot{r}}{r}+ (D-2)\frac{\dot{R}}{R}]\Phi_H\} +[-\frac{k_R^{(0)}}{R}
 +\frac{D}{k_R^{(0)}}(\ddot{R}-\frac{\dot{R}^2}{R}) ] V_s \\
&-2[d\frac{\dot{r}}{r}(\frac{k_R^{(0)}}{R})^2+D\frac{\dot{R}}{R}(\frac{k_r^{(0)}}{r})^2]\tilde{\Phi}_G,
\end{split}
\end{equation}
where $n = d+D, p=\rho/n$, a dot denotes $d/dt$,  $k_r^{(0)}$ and $k_R^{(0)}$ are the
wave-numbers in the outer and inner spaces, respectively, and $r= r_0 \tau^{-1/3}, \
R=R_0 \tau^{1/3}, \ \tau = t_0 - t$, and $t_0$ denotes the epoch of $r \rightarrow \infty$
 and $R = 0$.

Equations for $\dot{v}_s$ and $\dot{V}_s$ are obtained from  $\delta T^A_{i; A} = 0$ and
$\delta T^A_{a; A} = 0$, respectively, as  
\begin{equation}
  \label{eq:a2}
 \begin{split} 
\dot{v}_s &+ (\frac{\dot{r}}{r}-\frac{D}{n}\frac{\dot{R}}{R}) v_s = -\frac{D}{n}
\frac{k_r^{(0)}}{k_R^{(0)}} \frac{\dot{R}}{r}V_s + \frac{\epsilon_m}{n+1}\frac{k_r^{(0)}}{r}\\
&- \frac{k_r^{(0)}}{r} [(d-2) \Phi_h + D\Phi_H + 2(\frac{k_R^{(0)}}{R})^2 \tilde{\Phi}_G
+\frac{n+1}{n}D\frac{\dot{R}}{R}\Phi_7],
\end{split}
\end{equation}
and
\begin{equation}
  \label{eq:a3}
 \begin{split} 
\dot{V}_s &+ (\frac{\dot{R}}{R}-\frac{d}{n}\frac{\dot{r}}{r}) V_s = -\frac{d}{n}
\frac{k_R^{(0)}}{k_r^{(0)}} \frac{\dot{r}}{R}v_s + \frac{\epsilon_m}{n+1}\frac{k_R^{(0)}}{R}\\
&- \frac{k_R^{(0)}}{R} [(D-2) \Phi_H + d\Phi_h + 2(\frac{k_r^{(0)}}{r})^2 \tilde{\Phi}_G
-\frac{n+1}{n} d\frac{\dot{r}}{r}\Phi_7].
\end{split}
\end{equation}
From the latter two equations we obtain 
\begin{equation}
  \label{eq:a4}
   \begin{split} 
(\frac{r}{k_r^{(0)}} v_s - \frac{R}{k_R^{(0)}} V_s)^.  &= 2(\Phi_h - \Phi_H) 
+\frac{1}{n}(d\frac{\dot{r}}{r}+D\frac{\dot{R}}{R})(\frac{r}{k_r^{(0)}} v_s - 
\frac{R}{k_R^{(0)}} V_s)\\
&-2[(\frac{k_r^{(0)}}{r})^2 - (\frac{k_R^{(0)}}{R})^2] \tilde{\Phi}_G
-\frac{n+1}{n}(d\frac{\dot{r}}{r}+D\frac{\dot{R}}{R})\Phi_7,
\end{split}
\end{equation}
and integrating this above equation,  we have
\begin{equation}
  \label{eq:a5}
   \begin{split} 
\frac{r}{k_r^{(0)}} v_s - \frac{R}{k_R^{(0)}} V_s  &= -\tau^{1/9} \int d\tau \ \tau^{-1/9}
\{2(\Phi_h - \Phi_H) -2[(\frac{k_r^{(0)}}{r})^2 - (\frac{k_R^{(0)}}{R})^2] \tilde{\Phi}_G\\
&-\frac{n+1}{n}(d\frac{\dot{r}}{r}+D\frac{\dot{R}}{R})\Phi_7 \} \ \equiv \ A_0(\Phi_h, \Phi_H,
\tilde{\Phi}_G, \Phi_7).
\end{split}
\end{equation}
From Eqs. (\ref{eq:a1}) and  (\ref{eq:a5}), we obtain
\begin{equation}
  \label{eq:a6}
\frac{n}{n+1} \dot{\epsilon}_m  = \frac{1}{n+1}(d\frac{\dot{r}}{r}+D\frac{\dot{R}}{R})
\epsilon_m - A_1 + C_h v_s + C_H V_s,
\end{equation}
and from Eq. (\ref{eq:a5})
\begin{equation}
  \label{eq:a7}
V_s = \frac{k_R^{(0)}}{R} (\frac{r}{k_r^{(0)}} v_s - A_0),
\end{equation}
where
\begin{equation}
  \label{eq:a8}
     \begin{split} 
A_1 &\equiv d\{\dot{\Phi}_h +[D\frac{\dot{R}}{R}+(d-2)\frac{\dot{r}}{r}]\Phi_h\} \\
&+D\{\dot{\Phi}_H +[d\frac{\dot{r}}{r}+(D-2)\frac{\dot{R}}{R}]\Phi_H\}
+2 [d \frac{\dot{r}}{r}(\frac{k_R^{(0)}}{R})^2 +D \frac{\dot{R}}{R}(\frac{k_r^{(0)}}{r})^2] \tilde{\Phi}_G, 
\end{split}
\end{equation}
\begin{equation}
  \label{eq:a10}
C_h \equiv -\frac{k_r^{(0)}}{r} + \frac{d}{k_r^{(0)}} (\ddot{r} -\dot{r}^2/r), 
\end{equation}
\begin{equation}
  \label{eq:a11}
C_H \equiv -\frac{k_R^{(0)}}{R} + \frac{D}{k_R^{(0)}} (\ddot{R} -\dot{R}^2/R).
\end{equation}

Next, differentiating Eq. (\ref{eq:a6}) with respect to $t$ and eliminating  $v_s$ and $\dot{v}_s$
using Eqs. (\ref{eq:a2}) and (\ref{eq:a6}), we obtain the following equations for 
$\epsilon_m$ and $v_s$ 
\begin{equation}
  \label{eq:a12}
\ddot{\epsilon}_m + D_0 \dot{\epsilon}_m + D_1 \epsilon_m = \frac{n+1}{n} E
\end{equation}
and
\begin{equation}
  \label{eq:a13}
v_s = \frac{1}{\tilde{C}} \{ \frac{1}{n+1} [n\dot{\epsilon}_m -(d\frac{\dot{r}}{r}+
D\frac{\dot{R}}{R}) \epsilon_m] +A_1 + C_H \frac{k_R^{(0)}}{R} A_0\},
\end{equation}
where 
\begin{equation}
  \label{eq:a14}
\tilde{C} \equiv C_h + \frac{k_R^{(0)}}{k_r^{(0)}} \frac{r}{R} C_H,
\end{equation}
\begin{equation}
  \label{eq:a15}
D_0 \equiv \frac{D}{n} (\frac{\dot{r}}{r} -\frac{\dot{R}}{R}) - \frac{\dot{\tilde{C}}}{\tilde{C}}, 
\end{equation}
\begin{equation}
  \label{eq:a16}
D_1\equiv -\frac{1}{n} (d\frac{\dot{r}}{r} +D\frac{\dot{R}}{R})^.  -\frac{1}{n} 
(d\frac{\dot{r}}{r} +D\frac{\dot{R}}{R}) (\frac{\dot{r}}{r} -\frac{\dot{\tilde{C}}}{\tilde{C}})
- \frac{k_r^{(0)}}{r} \frac{\tilde{C}}{n},
\end{equation}
\begin{equation}
  \label{eq:a9}
A_2 \equiv (d-2) \Phi_h +D\Phi_H +2(\frac{k_R^{(0)}}{R})^2 \tilde{\Phi}_G + \frac{n+1}{n} D 
\frac{\dot{R}}{R} \Phi_7.
\end{equation}
\begin{equation}
  \label{eq:a17}
E \equiv - (A_1 + \frac{k_R^{(0)}}{R} C_H A_0)(\frac{\dot{r}}{r} -\frac{\dot{\tilde{C}}}{\tilde{C}})
- \dot{A}_1 - k_R^{(0)} (\frac{C_H}{R} A_0)^{.} + (\frac{D}{n} \frac{\dot{R}}{R} A_0 - A_2)
\frac{k_r^{(0)}}{r} \tilde{C},
\end{equation}

\subsection{Outside the horizons}

At epoch $t_{dec}$ when the outer space and the inner space decouple, $\tau$ is assumed 
to be so small that 
\begin{equation}
  \label{eq:a18}
x \equiv \frac{3}{4r_0}  k_r^{(0)}  \tau^{4/3} \ll 1
\end{equation}
and
\begin{equation}
  \label{eq:a19}
y \equiv \frac{3}{2R_0}  k_R^{(0)}  \tau^{2/3} \ll 1.
\end{equation}
Under these conditions the perturbations with wave-numbers $k_r^{(0)}$ and $k_R^{(0)}$
are outside the horizons in the outer and inner spaces, and we have the relations
\begin{equation}
  \label{eq:c3}
\Phi_h = (\tau/\tau_i)^{-8/3} (\Phi_h)_i + \Delta \Phi_h, 
\end{equation}
\begin{equation}
  \label{eq:c4}
\Phi_H = (\tau/\tau_i)^{-4/3} (\Phi_H)_i + \Delta \Phi_H, 
\end{equation}
where $\Delta \Phi_h$ and $\Delta \Phi_H$ consist of higher-order terms  
$O (x^2)$ and $O (y^2)$ with respect to $x$ and $y$, respectively, which are shown 
in Appendix A, and $\tau_i$ is an arbitrary epoch at the stage when Eqs.(\ref{eq:a18}) and
 (\ref{eq:a19}) are satisfied. 

Now let us derive $\epsilon_m, v_s$ and $V_s$ corresponding to the above curvature
perturbations,  neglecting higher-order terms, such as $\Delta \Phi_h, \Delta \Phi_H,
 \tilde{\Phi}_G$ and $\Phi_6$. Here we must pay attention to $\Phi_7$. 
 
Substituting Eqs.(\ref{eq:c3}) and (\ref{eq:c4}) into Eqs.(\ref{eq:a14}) $\sim$ 
(\ref{eq:a17}), we obtain
\begin{equation}
  \label{eq:a21}
     \begin{split} 
C_h &= - \frac{k_r^{(0)}}{r_0} \tau^{1/3} + \frac{r_0}{k_r^{(0)}} \tau^{-7/3}, \quad 
C_H = - \frac{k_R^{(0)}}{R_0} \tau^{-1/3} - \frac{2R_0}{k_R^{(0)}} \tau^{-5/3},  \\ 
\tilde{C} &\simeq - \frac{r_0}{k_r^{(0)}} \tau^{-7/3}, \quad \dot{\tilde{C}}/\tilde{C} = 
-{\tilde{C}}'/\tilde{C} = 7/(3\tau), \quad \Phi_7 \simeq 3\tau (\Phi_h + \Phi_H)
\end{split}
\end{equation}
\begin{equation}
  \label{eq:a22}
D_0 \simeq - 17/(9\tau), \quad D_1 \simeq 0,
\end{equation}
Moreover,
\begin{equation}
  \label{eq:a23}
     \begin{split} 
A_0 & \simeq 3\tau(\Phi_h + \Phi_H), \quad A_1 = (3/\tau)(\Phi_h + 2\Phi_H),  \\
A_2 & \simeq - \frac{1}{3} (17\Phi_h +2\Phi_H), \quad  E \simeq 0 .
\end{split}
\end{equation}
From Eqs. (\ref{eq:a12}), (\ref{eq:a13}) and (\ref{eq:a7}), we obtain in the lowest-order
\begin{equation}
  \label{eq:a24}
\epsilon_m = 0,
\end{equation}
\begin{equation}
  \label{eq:a25}
v_s =  4 x \ \Phi_h ,
\end{equation}
and 
\begin{equation}
  \label{eq:a26}
V_s =  - 2 y \ \Phi_H.
\end{equation}
So, $\epsilon_m$ is of higher-orders ($\sim O(x^2), O(y^2)$).
Here $\Phi_h$ and $\Phi_H$ are independent, because $\Phi_{hi}$ and $\Phi_{Hi}$ are
given arbitrarily. 
 
\subsection{Inside the horizons} 
 
At earlier epochs of $\tau \gg \tau_{dec}$, $x$ and $y$ are comparable with $1$ or 
larger than $1$.  It was shown in Sect. 3 of [5] that                           
in the case of $x \gg 1$ and $y\gg 1$ under the condition 
\begin{equation}
  \label{eq:a26z}
 \mu/x (\equiv [(k_R^{(0)}/{R(t)})/(k_r^{(0)}/{r(t)})]^2) \ll 1,
\end{equation}
the perturbations show
 wavy behaviors, depending on the wave-number $k_r^{(0)}$ (in the outer space) as 
 $exp(i\omega x)$,  where $\omega$ is a constant.
 In the case of  $y \gg 1$ and $x\gg 1$ under the
 condition $\mu/x \gg 1$, the perturbations show
 wavy behaviors, depending on the wave-number $k_R^{(0)}$ (in the inner space) as 
 $exp(i\omega y)$.                                
In the former case ($\mu/x \ll 1$) the waves (depending on $k_r^{(0)}$) appear mainly 
in the outer space but do not appear in the inner space. In the latter case ($\mu/x \gg 1$), 
on the other hand, the waves 
(depending on  $k_R^{(0)}$) appear mainly in the inner space but do not appear in the
 outer space.                                                         

At the stage of $x \gg 1$ and $y \gg 1$, the perturbations are inside the horizons and
can be created by quantum fluctuations in both  the outer space and the inner space.
For these perturbations we assume that the values of $k_r^{(0)}$ and $k_R^{(0)}$ are 
smoothly distributed around the average values ($\bar{k}_r^{(0)}$ and $\bar{k}_R^{(0)}$). 
Here we pay attention to the perturbations with $x \gg 1, \bar{y} (\equiv (3/2R_0) 
\bar{k}_R^{(0)}) \gg 1$ and $\bar{\mu}/x \equiv
[(\bar{k}_R^{(0)}/R(t)]/[k_r^{(0)}/r(t)]^2 \ll 1$. In this case, the perturbations with large
${k}_r^{(0)}$ have wavy behaviors in the outer space and the perturbations with
${k}_R^{(0)} \approx \bar{k}_R^{(0)}$ in the inner space are negligible. In the following 
we study their wavy behaviors
proportional to $\exp (i\omega x) $. This is because such perturbations will survive
and may be connected 
with the present observational information through the CMB radiation,
after the decoupling of the outer space from the inner space.
The other perturbations including the components ($\propto \exp i \omega y$) 
in the inner space will be disturbed or erased when the inner space is decoupled and disapear.

For perturbations with $x \gg 1, \bar{y} \gg 1$ and $\bar{\mu}/x \ll 1$, 
it is found from Eqs. (\ref{eq:a10}),
(\ref{eq:a11}),  (\ref{eq:a14}), (\ref{eq:a15}) and (\ref{eq:a16}) that 
\begin{equation}
  \label{eq:a26a}
     \begin{split} 
C_h & \simeq -k_r^{(0)}/r,  \quad |C_H | \simeq \bar{k}_R^{(0)}/R \ll |C_h|, \\
\tilde{C} & \simeq C_h,  \quad \dot{\tilde{C}}/\tilde{C} \simeq -\dot{r}/r, 
\end{split}
\end{equation}
and 
\begin{equation}
  \label{eq:a26b}
D_0 \simeq \frac{7}{9\tau},  \quad D_1 \simeq (\frac{k_r^{(0)}}{r})^2 \frac{1}{9}(1+\frac{15}{16} x^{-2}).
\end{equation}
Here let us put $\epsilon_m$ and curvature perturbations $\Phi_h$ and $\Phi_H$ as
\begin{equation}
  \label{eq:a26c}
{\epsilon_m = {\epsilon_{m0}\exp i\omega x, \quad \Phi_h = \Phi_{h0} \exp i\omega x}, \quad
\Phi_H = \Phi_{H0} \exp i\omega x}.
\end{equation}
Then we obtain for $x \gg 1$
\begin{equation}
  \label{eq:a27}
     \begin{split} 
\ddot{\epsilon}_m &+D_0 \dot{\epsilon}_m +D_1 \epsilon_m \\  
&\simeq (k_r^{(0)}/r)^2 [(\frac{1}{9}-\omega^2)\epsilon_{m0} +i \omega (2\epsilon_{m0,x} 
-\frac{1}{3}\epsilon_{m0}/x)\\
&+ \epsilon_{m0,xx} -\frac{1}{3} \epsilon_{m0,x}/x + \frac{5}{48} \epsilon_{m0}/x^2] \
\exp i\omega x \\
&= (\frac{k_r^{(0)}}{r^2})^2 (\frac{1}{9} -\omega^2) \ \epsilon_{m0} \ \exp i\omega x
 \ [1 + O(1/x)] \quad {\rm for} \ \omega \ne 1/3.
\end{split}
\end{equation}
On the other hand, we get from Eqs. (\ref{eq:a5}) and  (\ref{eq:a8}) 
\begin{equation}
  \label{eq:a28}
     \begin{split}   
A_0 &= - \frac{3\tau/2}{i\omega x} \ (\Phi_h-\Phi_H)  
 \ [1 + O(1/x)],\\
A_1 &= - \frac{k_r^{(0)}}{r} i\omega (d\Phi_h + D\Phi_H) \ [1 + O(1/x)], \\
\dot{A}_1 &= - {A_1}'  = - (\frac{k_r^{(0)}}{r})^2 \omega^2 (d\Phi_h + D\Phi_H)
\ [1 + O(1/x)], \\
A_2 &= (\Phi_h + 6 \Phi_H)  \ [1 + O(1/x)].
\end{split}
\end{equation}
Here it is noticed that $\tilde{\Phi}_G$ is of higher-order with
respect to $1/x \ (\ll 1)$ and is neglected. From Eqs. (78) and (81) of [5], it is found that 
$\Phi_6 = (2, 6) \tau \Phi_h$ for $\omega = 1, 1/3$, respectively, so that $\Phi_7 = 0$
for $\omega = 1, 1/3$.  Moreover, using the condition 
$\bar{\mu}/x \ll 1$, we find that 
\begin{equation}
  \label{eq:a28a}
\frac{\bar{k}_R^{(0)}}{R} |C_H A_0| \ll |A_1|.
\end{equation}
Then from Eqs. (\ref{eq:a17}) and (\ref{eq:a28}), we obtain
\begin{equation}
  \label{eq:a30}
E \simeq (\frac{k_r^{(0)}}{r})^2 [(1+3 \omega^2) \Phi_{h0} + 6 (1 + \omega^2)
\Phi_{H0}] \exp i\omega x \ [1 + O(1/x)]. 
\end{equation}
So it is found from Eqs. (\ref{eq:a12}), (\ref{eq:a27})  and  (\ref{eq:a30})  that  
\begin{equation}
  \label{eq:a31}
\epsilon_{m0} \simeq  (\frac{1}{9} - \omega^2)^{-1} \frac{10}{9} 
[(1+ 3 \omega^2) \Phi_{h0}
+ 6 (1+\omega^2) \Phi_{H0}] \ [1 + O(1/x)] \quad {\rm for} \ \omega \ne 1/3 
\end{equation}
and
\begin{equation}
  \label{eq:a32}
\epsilon_{m0} \simeq - i x \frac{40}{27} ( \Phi_{h0} + 5 \Phi_{H0})  \quad {\rm for} 
\ \omega = 1/3. 
\end{equation}

In Sect. 3 of [5], it was found that the approximate wavy solutions of equations for curvature
 perturbations  are given only for $\omega = 1$ and $1/3$, and in these cases the 
 solutions have the following relations
\begin{equation}
  \label{eq:a33}
  \Phi_{H0} = - \frac{1}{3} \Phi_{h0} \quad {\rm for} \ \omega = 1, 
\end{equation}
and 
\begin{equation}
  \label{eq:a34}
  \Phi_{H0} = \Phi_{h0} \quad {\rm for} \ \omega = 1/3.
\end{equation}
So we have the following expressions for $\epsilon_{m0}$
\begin{equation}
  \label{eq:a35}
\epsilon_{m0}  = \ O(1/x) \ \Phi_{h0}  \quad \ \omega = 1, 
\end{equation}
and 
\begin{equation}
  \label{eq:a36}
\epsilon_{m0} \simeq  - \frac{80}{9} ix \Phi_{h0}  \quad {\rm for} \ \omega = 1/3. 
\end{equation}
%

\section{Quantum fluctuations}         

  In Sect. 2 and the previous paper [5], the perturbations were classified into three modes. 
In this paper 
we treat only their scalar and tensor modes and consider the perturbations created by
the quantum effect in the comparably later stage of the $10$-dimensional universe 
which is associated with the inflating 
outer space and collapsing inner space. Here Weinberg's procedure is used for the 
quantization \citep{wein}. 
  
\subsection{The scalar mode}

  At the stage of $x \gg 1$ and $\bar{y} \gg 1$, the length of perturbations in 
the outer space can be smaller than the horizon size and
they may be caused by the quantum effect, while at the later stage of $x < 1$ the length of
perturbations is larger than the horizon size and they  are frozen. 
So we should first consider the 
quantum fluctuations at earlier epochs of $x \gg 1$ and $\bar{y} \gg 1$. Additionally, 
moreover, we assume that 
$\bar{\mu}/x \ll 1$, corresponding to the perturbations in the inner space with the 
average value ($\bar{k}_R^{(0)}$), which was described in Sect. 3. Then these 
perturbations appear mainly in the outer space, and hence  
we can treat the perturbations, as if they are in the $4$-dimensional space-time 
(consisting of the time $t$ and the outer space).

The energy density perturbation $\epsilon_m$ is expressed by Eq. (\ref{eq:a12}) in 
connection with gravitational perturbations. This equation is also derived from the action 
principle as
\begin{equation}
  \label{eq:b1}
I = \int dt d^3 \bm{x}  ({\cal L}_\epsilon + {\cal L}_g),
\end{equation}
where ${\cal L}_\epsilon$ and ${\cal L}_g$ are the fluidal and gravitational parts of the total
 Lagrangian, and  $\bm{x}$ is the coordinate in the outer space.
  Here ${\cal L}_\epsilon$ can be derived from Eq. (\ref{eq:a12}) as follows.
 At the stage of $x \gg 1, \bar{y} \gg 1$ and $\bar{\mu}/x \ll 1$, we have $D_0 $ and 
 $D_1$ in Eq. (\ref{eq:a26b}), and then the fluidal part in the equation of motion is
 derived using the following Lagrangian
\begin{equation}
  \label{eq:b2}
{\cal L}_\epsilon = \frac{1}{2} r^{7/3} [(\frac{\partial \epsilon_m}{\partial t})^2 + 
(\frac{1}{3r})^2 (\frac{\partial \epsilon_m}{\partial \bm{x}})^2],
\end{equation}
where $r = r_0 (t_0-t)^{-1/3}$,  and $\partial \epsilon_m/\partial \bm{x} = 
i \bm{k}_r^{(0)} \epsilon_m$ for $\epsilon_m \propto \exp (i {\bm{k}_r^{(0)}} \bm{x})$. 

On the other hand,  $\epsilon_m$ can be expanded as  
\begin{equation}
  \label{eq:b3}
\epsilon_m  (\bm{x},t) = \int d \bm{k}_r^{(0)} [\epsilon_m (k_r^{(0)}, t) \exp (i\bm{k}_r^{(0)} \bm{x})
\ \alpha(\bm{k}_r^{(0)}) +\epsilon_m^* (k_r^{(0)}, t) \exp (-i\bm{k}_r^{(0)} \bm{x}) \
\alpha^* (\bm{k}_r^{(0)})],
\end{equation}
and $\Phi_h$ and $\Phi_H$ also can be written as
\begin{equation}
  \label{eq:b4}
 \begin{split} 
\Phi_h  (\bm{x},t) &= \int d \bm{k}_r^{(0)} [\Phi_h (k_r^{(0)}, t) \exp (i\bm{k}_r^{(0)} \bm{x}) 
\ \alpha(\bm{k}_r^{(0)}) +\Phi_h^* (k_r^{(0)}, t) \exp (-i\bm{k}_r^{(0)} \bm{x}) \
\alpha^* (\bm{k}_r^{(0)})],\\
\Phi_H  (\bm{x},t) &= \int d \bm{k}_r^{(0)} [\Phi_H (k_r^{(0)}, t) \exp (i\bm{k}_r^{(0)} \bm{x})
\ \alpha(\bm{k}_r^{(0)}) +\Phi_H^* (k_r^{(0)}, t) \exp (-i\bm{k}_r^{(0)} \bm{x}) \
\alpha^* (\bm{k}_r^{(0)})],
 \end{split} 
\end{equation}
where the reality of these fields requires to take the above forms. The interaction of the 
photon field with the gravitational field makes the commutation relation of 
$\alpha(\bm{k}_r^{(0)})$ and $\alpha^* (\bm{k}_r^{(0)})$ complicated, but they become
simple at very early times.\citep{wein} 

In many cases when quantum fluctuations have so far been treated in a system of a scalar
 (inflaton) field and the gravitational field, the quantization of the scalar field is first 
tried.\citep{wein}  In the present case also when we consider a system of a photon 
scalar field and the gravitational field,  we try first the quantization of the photon scalar 
field in the following. 

The canonical conjugate to $\epsilon_m  (\bm{x},t)$ is then
\begin{equation}
  \label{eq:b5}
\pi_m (\bm{x},t) = \partial {\cal L}_\epsilon / \partial (\frac{\partial \epsilon_m}{\partial t})
= r^{7/3} \frac{\partial \epsilon_m}{\partial t}.
\end{equation}
The commutator of   $\epsilon_m$ and  $\pi_m$ is
\begin{equation}
  \label{eq:b6}
[\epsilon_m (\bm{x},t),  \epsilon_m (\bm{y},t)] = 0, \quad  [\epsilon_m (\bm{x},t),  
\partial \epsilon_m (\bm{y},t)/\partial t] = i r^{-7/3} \delta^3 (\bm{x} - \bm{y}).
\end{equation}
These commutation relations imply that $\alpha (\bm{k})$ and $\alpha^\star (\bm{k})$
behave as conventionally normalized annihilation and creation operators
\begin{equation}
  \label{eq:b7}
[\alpha (\bm{k}), \alpha (\bm{k}')] = 0, \ \quad [\alpha (\bm{k}), \alpha^* (\bm{k}')] =
\delta^3 (\bm{k} - \bm{k}'),
\end{equation}
when $\epsilon_m (k_r^{(0)}, t)$ is normalized at $r \rightarrow 0$  as 
\begin{equation}
  \label{eq:b8}
\epsilon_m (k_r^{(0)}, t)  \propto [r(t)]^{-2/3} [k_r^{(0)}]^{-1/2} \exp (i \omega k_r^{(0)} 
\int^t_{t_*} \frac{dt'}{r(t')}),
\end{equation}
where $t_*$ is arbitrary and $\omega$ is a constant ($= 1$ or $1/3$).
This expression of $\epsilon_m (k_r^{(0)}, t)$ is used as the initial condition for created
fields of energy density $\epsilon_m$. 
Here we choose the quantum state during the inflation of the outer space under the simple
 assumption that the state of the universe is the vacuum state $|0 \rangle$, defined so that
\begin{equation}
  \label{eq:b9}
\alpha (\bm{k}) |0\rangle = 0  \quad {\rm and} \quad \langle 0|0 \rangle = 1. 
\end{equation}
This corresponds to the Bunch-Davies vacuum\citep{bunch} in the outer space within the 
$10$-dimensional universe. As described in Sect. 3, $\epsilon_m$ and the curvature 
perturbations as quantum fluctuations are proportional each other. So the behavior of 
$\epsilon_m$ in Eq. (\ref{eq:b8}) is common to that of $\Phi_h$ and $\Phi_H$, and,
using Eqs. (\ref{eq:a33}) - (\ref{eq:a36}), we obtain 
\begin{equation}
  \label{eq:b10}
\Phi_H (k_r^{(0)}, t) = - \frac{1}{3} \Phi_h (k_r^{(0)}, t) 
\propto x \ [r(t)]^{-2/3} [k_r^{(0)}]^{-1/2} \exp (i \omega k_r^{(0)} \int^t_{t_*} \frac{dt'}{r(t')})
\end{equation}
for  $\omega = 1$, and
\begin{equation}
  \label{eq:b10a}
\Phi_H (k_r^{(0)}, t) = \Phi_h (k_r^{(0)}, t) 
\propto x^{-1} \ [r(t)]^{-2/3} [k_r^{(0)}]^{-1/2} \exp (i \omega k_r^{(0)} \int^t_{t_*} \frac{dt'}{r(t')})
\end{equation}
for $\omega = 1/3$.

\subsection{The tensor mode}

In the tensor mode, there are two types (ST) and (TS), as described in Sect. 2 and 
[5]. (ST) has the 
$3$-dimensional scalar and the $6$-dimensional tensor, while (TS) has the $3$-dimensional
tensor and the $6$-dimensional scalar. Here we take up (TS) with the amplitude $h_T^{(2)}$,
and neglect (ST) with the amplitude $H_T^{(2)}$, which may not be connected with the 
observation in the $3$-dimensional outer space, after the decoupling of the inner space.

In [5], we studied the behavior of tensor perturbations $h_T^{(2)}$. They satisfy
\begin{equation}
  \label{eq:b11}
\ddot{h}_T^{(2)} +(d\frac{\dot{r}}{r}+D\frac{\dot{R}}{R}) \dot{h}_T^{(2)}+ 
[(\frac{k_r^{(2)}}{r})^2+ (\frac{k_R^{(0)}}{R})^2] h_T^{(2)} = 0.
\end{equation}
Here we consider the case of
\begin{equation}
  \label{eq:b12}
(\frac{\bar{k}_R^{(0)}}{R})/(\frac{k_r^{(2)}}{r}) \ll 1,
\end{equation}
where $k_r^{(2)}$ and $k_R^{(0)}$ are the wave-numbers in the outer and inner spaces,
 respectively, and $\bar{k}_R^{(0)}$ is the average wave-number in the inner space.
 Then  we have
\begin{equation}
  \label{eq:b13}
\ddot{h}_T^{(2)} - 3\frac{\dot{r}}{r} \dot{h}_T^{(2)}+ (\frac{k_r^{(2)}}{r})^2 h_T^{(2)} = 0,
\end{equation}
where we used the relation $R \propto 1/r \propto \tau^{1/3}$ and $d = D/2= 3$. 
This equation can be also derived from the action principle as
\begin{equation}
  \label{eq:b14}
I = \int dt d^3 x \ {\cal L}_t, 
\end{equation}
where 
\begin{equation}
  \label{eq:b15}
{\cal L}_t = \frac{1}{2} r^{-3} [(\frac{\partial h_T^{(2)}}{\partial t})^2 + \frac{1}{r^2} 
(\frac{\partial h_T^{(2)}}{\partial x^i})^2]
\end{equation}
for $\partial h_T^{(2)} \propto \exp i \bm{k}_r^{(2)} \bm{x}$.

On the other hand, the amplitude $\partial h_T^{(2)}$ takes the form
\begin{equation}
  \label{eq:b16}
 h_T^{(2)} (\bm{x},t) = \int d \bm{k}_r^{(2)} \ [h_T^{(2)} (k_r^{(2)}, t) \exp (i\bm{k}_r^{(2)} \bm{x})
\ \alpha(\bm{k}_r^{(2)}) + h_T^{(2)*} (k_r^{(2)}, t) \exp (-i\bm{k}_r^{(2)} \bm{x}) \
\alpha^* (\bm{k}_r^{(2)})],
\end{equation}
and the canonical conjugate  to $h_T^{(2)} (\bm{x},t)$ is then
\begin{equation}
  \label{eq:b17}
\pi_T (\bm{x},t) = \partial {\cal L}_t /(\partial h_T^{(2)}/{\partial t})
= r^{-3} \frac{\partial h_T^{(2)}}{\partial t}.
\end{equation}
The commutator of $h_T^{(2)}$ and $\pi_T$ is
\begin{equation}
  \label{eq:b18}
[h_T^{(2)} (\bm{x},t),  h_T^{(2)} (\bm{y},t)] = 0, \  [h_T^{(2)} (\bm{x},t),  
\partial h_T^{(2)} (\bm{y},t)/\partial t] = i r^{3} \delta^3 (\bm{x} - \bm{y}).
\end{equation}
These commutation relations imply that $\alpha (\bm{k})$ and $\alpha^* (\bm{k})$
behave as conventionally normalized annihilation and creation operators, in the same way as
Eq. (\ref{eq:b6}), when $h_T^{(2)} (k_r^{(2)}, t)$ is normalized at $r \rightarrow 0$ as
\begin{equation}
  \label{eq:b19}
h_T^{(2)} (k_r^{(2)}, t)  \propto [r(t)]^2 [k_r^{(2)}]^{-1/2} \exp (i k_r^{(2)} \int^t_{t_*} 
\frac{dt'}{r(t')}).
\end{equation}
This expression of $h_T^{(2)} (k_r^{(2)}, t)$ is used as the initial condition of created
fields in the tensor mode  $h_T^{(2)}$. 
Here we choose the quantum state during the inflation of the outer space, so that 
the state of the universe may satisfy the relation in Eq. (\ref{eq:b9}).
      
\section{Spectra of fluctuations and their comparison with CMB observation}
\subsection{The scalar mode}

 The information about the perturbations which are created by the quantum fluctuations inside 
 the horizon can be used to make an initial condition for the evolution of perturbations 
 which re-enter the horizon after the long inflation. For this purpose, we use the quantities 
 which are conserved outside the horizon. In the $4$-dimensional universe with 
 $3$-dimensional space-section, we have a gauge-invariant curvature perturbation, 
 represented as
\begin{equation}
  \label{eq:c1}
{\cal R}_4 \equiv \Phi_H,
\end{equation}
which is a conserved quantity.\citep{bar} In the $10$-dimensional universe, 
on the other hand, we have the following two independent similar 
quantities as the candidates 
\begin{equation}
  \label{eq:c2}
{\cal R}_h \equiv (\tau/\tau_{dec})^{8/3} \Phi_h \quad {\rm and} \quad {\cal R}_H \equiv
 (\tau/\tau_{dec})^{4/3} \Phi_H,
\end{equation}
where $\tau_{dec}$ represents the epoch when the inner space decouples from the 
outer space. For $x (\equiv (3/4r_0) k_r^{(0)} \tau^{4/3}) < 1$ and $y (\equiv (3/2R_0) 
k_R^{(0)} \tau^{2/3}) < 1$,  ${\cal R}_h$ and  ${\cal R}_H$ are nearly constant, and so
these can be regarded as quantities conserved outside the horizon. 
   
As other candidates for conserved quantities, we may consider 
\begin{equation}
  \label{eq:c5}
{\cal R}_v (\equiv \frac{\dot{r}}{k_r^{(0)}} v_s) \quad {\rm and} \quad 
{\cal R}_V (\equiv \frac{\dot{R}} {k_R^{(0)}} V_s), 
\end{equation}
but, for $x \ll 1$, they are not independent of  ${\cal R}_h$ and ${\cal R}_H$,
\begin{equation}
  \label{eq:c6}
{\cal R}_v = -\frac{23}{32} \Phi_h + \frac{19}{4} \Phi_H,
\end{equation}
\begin{equation}
  \label{eq:c7}
{\cal R}_V = -\frac{11}{32} \Phi_h + \frac{13}{4} \Phi_H
\end{equation}
with respect to the main terms.  For $x \gg 1$ and $y \gg 1$, moreover, we find
that $v_s$ and $V_s$ are comparable with $\Phi_h$ and $\Phi_H$, respectively, and  
${\cal R}_v$ and ${\cal R}_V$ are $\sim v_s/x$ and $\sim V_s/y$, respectively, 
which are small, compared with $\Phi_h$ and $\Phi_H$. This means that the roles of 
${\cal R}_v$ and ${\cal R}_V$ are small, compared with those of ${\cal R}_h$ and 
${\cal R}_H$, respectively. In this paper, therefore, we adopt ${\cal R}_h$ and ${\cal R}_H$
as the conserved quantities in the $10$-dimensional universe. Neither of them, however,
is necessarily a conserved quantity which is directly connected at epoch 
$\tau_{dec}$ with $R_4$ in the $4$-dimensional universe.

Here we construct the $10$-dimensional gauge-invariant 
conserved quantity ${\cal R}_{10}$ using ${\cal R}_h$ and ${\cal R}_H$, by imposing the
following two conditions :

\noindent (1) ${\cal R}_{10} = {\cal R}_{4}$ at epoch ($\tau_{dec}$) of the decoupling 
of the outer space from the inner space, and

\noindent (2)  ${\cal R}_{10}$ is consistent with the spectral constraint given by the CMB
 observation.
 
As the first candidate of ${\cal R}_{10}$, we consider a linear combination of $\Phi_h$ and $\Phi_H$ as
\begin{equation}
  \label{eq:c8}
{\cal R}_{10} =  \lambda_0 {\cal R}_H + \lambda_1 {\cal R}_h,
\end{equation}
where constants $\lambda_0$ and  $\lambda_1$ are determined so as to satisfy
 the above two conditions (1) and (2).

At epoch $\tau$ when $x \ (= (3/4r_0) k_r^{(0)} \tau^{4/3}) \gg 1$ and $y 
(= (3/4r_0) k_R^{(0)} \tau^{2/3}) \gg 1$,
 $\Phi_h$ and $\Phi_H$
are created by quantum fluctuations and they are expressed using Eqs.(\ref{eq:b10}) and
(\ref{eq:b10a}) as
\begin{equation}
  \label{eq:c11}
     \begin{split} 
\Phi_h  &= \tau^{14/9} [k_r^{(0)}]^{1/2} \exp (ix) +\alpha \tau^{-10/9} [k_r^{(0)}]^{-3/2}
(4r_0/3)^2 \exp (ix/3),  \\
\Phi_H &= -\frac{1}{3} \tau^{14/9} [k_r^{(0)}]^{1/2} \exp (ix) +\alpha \tau^{-10/9}
 [k_r^{(0)}]^{-3/2} (4r_0/3)^2 \exp (ix/3),
\end{split}
\end{equation}
where we used $r \propto \tau^{-1/3}$ and $\alpha$ is an arbitrary constant.
Inserting Eq.(\ref{eq:c11}) into Eq.(\ref{eq:c2}), we obtain
\begin{equation}
  \label{eq:c12}
      \begin{split} 
{\cal R}_h  &=  {\tau_{dec}}^{-8/3} [k_r^{(0)}]^{1/2} [{\tau}^{38/9} \exp (ix) + \alpha 
{\tau}^{14/9} (\frac{4r_0}{3k_r^{(0)}})^2 \exp (ix/3)] \\
&=  [k_r^{(0)}]^{-8/3} {\tau_{dec}}^{-8/3} (4r_0 \ x/3)^{19/6} [\exp (ix) + 
\alpha x^{-2} \exp (ix/3)] ,
 \end{split}
\end{equation}
\begin{equation}
  \label{eq:c13}
    \begin{split} 
{\cal R}_H  &=  {\tau_{dec}}^{-4/3} [k_r^{(0)}]^{1/2} [-\frac{1}{3} {\tau}^{26/9}
\exp (ix) + \alpha {\tau}^{2/9} (\frac{4r_0}{3k_r^{(0)}})^2 \exp (ix/3)]  \\
&= [k_r^{(0)}]^{-5/3} {\tau_{dec}}^{-4/3} (4r_0 \ x/3)^{13/6} [-\frac{1}{3} \exp (ix) 
+ \alpha x^{-2} \exp (ix/3)] .
\end{split}
\end{equation}
As $x$ decreases and becomes smaller than $1$,
 the $x$ dependence of ${\cal R}_h$ and
${\cal R}_H$ changes from the wavy behavior ($\propto  \exp (ix) $ and $\exp (ix/3) $)
 to the constant ones. 
At epoch $\tau_{eq}$ with $x = 1$, we have therefore
\begin{equation}
  \label{eq:c12a}
{\cal R}_h (\tau_{eq}) =  \Xi \ (\zeta_h + \alpha \zeta'_h),
\end{equation}
\begin{equation}
  \label{eq:c13a}
{\cal R}_H (\tau_{eq}) = \Xi \ \frac{x_{dec}}{x} (-\frac{1}{3} \zeta_H + \alpha \zeta'_H),
\end{equation}
where
\begin{equation}
  \label{eq:c13b}
\Xi \equiv [k_r^{(0)}]^{-8/3} {\tau_{dec}}^{-8/3} (4r_0/3)^{19/6} ,
\end{equation}
where $x_{dec} \equiv (3/4r_0) k_r^{(0)} {\tau_{dec}}^{4/3}$, and $\zeta_h, \zeta'_h,
\zeta_H$ and $\zeta'_H$ are constants. The exact values of these constants 
 are determined by solving
 dynamical equations for $\Phi_h$ and $\Phi_H$ given in [5], but they are
estimated to be $\approx 1$, because ${\cal R}_h$ and  ${\cal R}_H$ are nearly 
constant for $x < 1$ and $y < 1$.

Now we assume that the CMB spectrum is determined at epoch $\tau_{eq}$ 
when $x = 1$ (indicating the horizon exit), and consider the $k_r^{(0)}$ 
dependence of ${\cal R}_{10}$ at this epoch. 
Here ${\cal R}_{10}$ at epoch $\tau_{eq}$ is expressed as
\begin{equation}
  \label{eq:c14}
{\cal R}_{10}  =  {\cal R}_{0} z^{-5/3} + {\cal R}_{1} z^{-8/3}
\end{equation}
around the observed wave-number $(k_r^{(0)})_{obs}$, where $z \equiv  
 k_r^{(0)}/(k_r^{(0)})_{obs}$,
\begin{equation}
  \label{eq:c15}
{\cal R}_{0} \equiv \lambda_0  {\rm Re}(-\frac{1}{3} \zeta_H + \alpha \zeta'_H) \
 (x_{dec} \ \Xi)_{z = 1} ,
\end{equation}
\begin{equation}
  \label{eq:c16}
{\cal R}_{1} \equiv  \lambda_1  {\rm Re}( \zeta_h + \alpha \zeta'_h) \ \Xi_{z = 1},
\end{equation}
where Re means the real part, and $\lambda_0$ and $\lambda_1$ are 
coefficients in  Eq.(\ref{eq:c8}). 

 The CMB observation shows that the  $k_r^{(0)}$ dependence of ${\cal R}_{10}$ is   
\begin{equation}
  \label{eq:c17}
z^{(-4+n_s)/2} = z^{-1.517}
\end{equation}
for $(k_r^{(0)})_{obs} = 0.002$ Mpc$^{-1}$, where $n_s = 0.966$ according to
the WMAP 7 year result\citep{kom}.     
The condition that  Eq.(\ref{eq:c14}) and Eq.(\ref{eq:c17}) should be consistent in the 
neighborhood of $z = 1$ is
\begin{equation}
  \label{eq:c18}
z^{-5/3} + \delta_1  z^{-8/3} = (1+\delta_1) z^{-1.517},
\end{equation}
where $\delta_1 \equiv {\cal R}_{1}/ {\cal R}_{0} = (x_{dec})^{-1} (\lambda_1/ \lambda_0) 
{\rm Re}(\zeta_h + \alpha \zeta'_h) /{\rm Re}(-\frac{1}{3}\zeta_H + \alpha \zeta'_H)$. 
 From the continuity of this equation
and its first derivative at $z = 1$, it is found that 
\begin{equation}
  \label{eq:c19}
\delta_1 = -0.130.
\end{equation}
That is, the observational spectrum (\ref{eq:c17}) can be reproduced when ${\cal R}_H$ is 
main and ${\cal R}_h$ is about $10 \%$ of the total ${\cal R}_{10}$. 

The above definition of ${\cal R}_{10}$ satisfies the condition of continuity of 
Eq.(\ref{eq:c18}) in the first derivative, but not in the second derivative.
In order to satisfy also the condition of continuity in the second derivative, we 
consider the second candidate of ${\cal R}_{10}$ at $\tau_{eq}$ expressed as
\begin{equation}
  \label{eq:c8a}
{\cal R}_{10} =  \lambda_0 {\cal R}_H + \lambda_1 {\cal R}_h 
+ \lambda_2 [{\cal R}_h]^2/{\cal R}_H,
\end{equation}
where constants $\lambda_0, \lambda_1$ and  $\lambda_2$ are determined so as to satisfy
 the above two conditions (1) and (2). Here  ${\cal R}_{10}$ at epoch $\tau_{eq}$ 
 is rewritten as
\begin{equation}
  \label{eq:c14a}
{\cal R}_{10}  =  {\cal R}_{0} z^{-5/3} + {\cal R}_{1} z^{-8/3} + {\cal R}_{2} z^{-11/3},
\end{equation}
where ${\cal R}_{0}$ and  ${\cal R}_{1}$ are defined by Eqs. (\ref{eq:c15}) and (\ref{eq:c16}),
and 
\begin{equation}
  \label{eq:c15a}
{\cal R}_{2} \equiv  \lambda_2 [{\rm Re}( \zeta_h + \alpha \zeta'_h)^2 
/{\rm Re}(-\frac{1}{3}\zeta_H + \alpha \zeta'_H)] [(x_{dec})^{-1} \Xi]_{z = 1}. 
\end{equation}
Then from the condition that Eqs.(\ref{eq:c14a}) and  (\ref{eq:c17}) should be
consistent in the neighborhood of $z = 1$, we have
\begin{equation}
  \label{eq:c18a}
z^{-5/3} + \delta_1 z^{-8/3} + \delta_2  z^{-11/3} = (1+\delta_1+ \delta_2) z^{-1.517},
\end{equation}
where $\delta_1 \equiv {\cal R}_{1}/ {\cal R}_{0} $ and $\delta_2 \equiv {\cal R}_{2}
/ {\cal R}_{0} $ .  From the continuity of this equation
and its first and second derivatives at $z = 1$, it is found that 
\begin{equation}
  \label{eq:c19a}
\delta_1 = -0.260 \quad {\rm and} \quad \delta_2 = 0.0696. 
\end{equation}

Now let us define the power spectrum of curvature perturbations as\citep{kom, felice}
\begin{equation}
  \label{eq:c20}
{\cal P}_s \equiv \frac{4\pi (k_r^{(0)})^3}{(2\pi)^3} |{\cal R}_{10}|^2.
\end{equation}
Then for $R_{10}, \delta_1$ and $\delta_2$ in Eqs.(\ref{eq:c8a}) and (\ref{eq:c19a}), 
\begin{equation}
  \label{eq:c22}
{\cal P}_s = \frac{4\pi [(k_r^{(0)})_{obs}]^3}{(2\pi)^3} |{\cal R}_{0}|^2 (1+\delta_1 + \delta_2)^2,
\end{equation}
The WMAP 7 year normalization\citep{kom} gives
\begin{equation}
  \label{eq:c23}
({\cal P}_s)_{obs} = 2.42 \times 10^{-9}
\end{equation}
on the scale $(k_r^{(0)})_{obs} = 0.002$ Mpc$^{-1}$. Then we obtain  
\begin{equation}
  \label{eq:c24}
{\cal R}_{0} = \lambda_0 |{\cal R}_H (\tau_{eq})| = \frac{\sqrt{2}\pi}{1+\delta_1 +\delta_2}
 {[{\cal P}_s /(k_r^{(0)})^3]_{obs}}^{1/2}.
\end{equation}
On the other hand, we have
\begin{equation}
  \label{eq:c24a}
     \begin{split} 
 \frac{\lambda_1}{\lambda_0} &=  \delta_1 \cdot
 x_{dec}   {\rm Re}(-\frac{1}{3}\zeta_H + \alpha \zeta'_H)/{\rm Re}( \zeta_h + \alpha \zeta'_h) , \\
\frac{\lambda_2}{\lambda_0} &= \delta_2 \cdot 
[x_{dec}  {\rm Re}(-\frac{1}{3}\zeta_H + \alpha \zeta'_H)/{\rm Re}( \zeta_h + \alpha \zeta'_h)]^2,
\end{split} 
\end{equation}
where $x_{dec} = (\tau_{dec}/\tau_{eq})^{4/3} = (r_{eq}/r_{dec})^4 \ll 1$, and the factor
${\rm Re}(-\frac{1}{3}\zeta_H + \alpha \zeta'_H)/{\rm Re}( \zeta_h + \alpha \zeta'_h)$ is of the 
order of $1$.  Since $\delta_2 \simeq (\delta_1)^2$, we have $\lambda_2/ \lambda_0 
\simeq (\lambda_1/\lambda_0)^2$.
 
It is concluded that ${\cal R}_{10}$ is consistent with the observed spectra of CMB radiation
under the condition of (\ref{eq:c24}) and (\ref{eq:c24a}).

Thus we could derive the condition that the parameters $\lambda_0, \lambda_1$ and  
$\lambda_2$ in $\mathcal{R}_{10}$ should satisfy for the consistency with the CMB observation.
From their ratios the role of the curvature perturbation in the inner space is found to be
larger than that in the outer space.
This condition and its consequeces are concerned with the condition at the earlier stage
and the initial condition of the universe, which should be expressed as a perturbation model
with the theoretical model parameters, and the above three parameters should be related
to the latter parameters. They may be influenced through $\mathcal{R}_{10}$ by the process of
 decoupling, which has not been discussed here, because its quantum-gravitational process 
cannot be treated at present. This situation in the observational aspect is compared with 
the situations in other inflation models, later in the subsection 5.3

\subsection{The tensor mode}

In the limit of $x \ (\equiv ({3}/{4r_0}) k_r^{(2)} \tau^{4/3}) \rightarrow 0$, the
gauge-invariant perturbation $h_T^{(2)}$ tends to $a + b \ln \tau$, as seen from
the analyses in Abbott et al.\citep{abb} and the previous paper [5], where $a$ and $b$ are 
constants. So, as the quantity ${\cal R}_t$ which is conserved outside the horizon, 
we adopt 
\begin{equation}
  \label{eq:c25}
{\cal R}_t \equiv h_T^{(2)} [a + b \ln \tau_{dec}]/[a + b\ln \tau],
\end{equation}
so that ${\cal R}_t$ leads to a constant in the limit of $x \rightarrow 0$. 

At the epoch $\tau_{eq}$ of $x = 1$, we have the relation
\begin{equation}
  \label{eq:c26}
\tau \propto [k_r^{(2)}]^{-3/4},
\end{equation}
so that $r^2 [k_r^{(2)}]^{-1/2} \propto \tau^{-2/3}  [k_r^{(2)}]^{-1/2} = const$,
and from Eq. (\ref{eq:b19})
\begin{equation}
  \label{eq:c27}
h_T^{(2)} (k_r^{(2)}, \tau_{eq}) = \lambda_t \cdot  \exp (ix),
\end{equation}
where $\lambda_t$ is a constant. Then it is found from Eq. (\ref{eq:c25})
that 
\begin{equation}
  \label{eq:c27a}
{\cal R}_t (\tau_{eq}) = \lambda_t [a + b \ln \tau_{dec}] 
[a - \frac{3}{4} b \ln k_r^{(2)} + const]^{-1} \exp (ix).
\end{equation}

As $x$ decreases and becomes $< 1$, the $x$ dependence of ${\cal R}_t$ changes from 
the wavy behavior to the stationary constant one. But the $k_r^{(2)}$ dependence does 
not change, so that the spectrum in the tensor mode have the form of 
\begin{equation}
  \label{eq:c28}
[a - \frac{3}{4} b \ln k_r^{(2)} + const]^{-1}.
\end{equation}
The corresponding power spectrum is
\begin{equation}
  \label{eq:c29}
{\cal P}_t \equiv \frac{4\pi (k_r^{(2)})^3}{(2\pi)^3} |{\cal R}_{t} (\tau_{eq})|^2.
\end{equation}

The amplitude of ${\cal R}_t (\tau_{eq})$ should be determined, corresponding to the 
observation, which has not been given yet. At present, we have the condition
$r \equiv {\cal P}_t / {\cal P}_s < 0.24$ for $k_r^{(0)} = k_r^{(2)} = 0.002$
Mpc$^{-1}$.\citep{kom}

\subsection{Comparison with the spectral analyses in other inflation models}
 
In the $4$-dimensional universe due to the Einstein theory, the quantity
conserved outside the horizon is uniquely defined using one of curvature 
perturbations.\citep{bar} 
But in hypothetical inflation models with inflaton scalar fields (including the non-minimal 
coupling with the Ricci scalar), the values of parameters such as slow-roll parameters 
($\epsilon, \ \eta$) and the number $N$ of inflationary e-folds,\citep{wein, ll} and
the coupling parameter $\xi$ in the scalar field equation\citep{fak, sal, 
kai, komhut}  are not unique. The observed spectral index $n_s \ (\approx 0.97)$ is,
therefore, obtained by adjusting the above parameters $\epsilon, \eta, N$ and $\xi$.

 In the $R + R^2$ modified gravitational theory, we have an inflation model 
 associated with the 
 de Sitter type solution which was derived first by Nariai and 
 Tomita\citep{nt} and rederived later by Starobinsky.\citep{st} 
  Mukhanov and Chibisov\citep{mc} derived the quantum 
 fluctuations generated at the de Sitter stage, and it was found that the spectral
 index $n_s$ of these fluctuations can be expressed as
\begin{equation}
  \label{eq:c30}
n_s - 1 = -1/[ 1 + \frac{1}{2} \ln (k_{obs}/a H)] = - 1/(1 + \frac{1}{2} N),
\end{equation}
where $k_{obs}$ is the observed wave-number, $N$ is the inflationary e-fold, and
$a$ and $H$ are the scale factor and the Hubble constant at the epoch when the
de Sitter expansion ends.  This 
number $N$ is determined to be $70$, so that we may have $n_s \simeq 0.97$ 
 (the observed value).

In the present case of a photon scalar field in the $10$-dimensional universe, 
 the inflation of the outer space is unique, because the scale factor 
$r$ of the outer space is $\propto \tau^{-1/3}$  (in the non-viscous case).
On the other hand, the conserved quantity is not unique, because there are 
two independent curvature 
perturbations $\Phi_h$ and $\Phi_H$ before the decoupling of the outer and inner spaces.
It is, therefore, a key point to determine how to combine them in this case, 
to derive ${\cal R}_{10}$ (connecting the two epochs outside the horizon). To obtain
the observed spectral index $n_s$, we made the examples of the combination of 
$\Phi_h$ and $\Phi_H$ as the conserved quantity, so that the theoretical    
spectral index $n_s$ may be consistent with the observed one.

\section{Concluding remarks}

In this paper I showed the possibility of deriving the observed fluctuation of CMB radiation
from the quantum fluctuations which appeared at the inflating stage of the outer space
in the $10$-dimensional universe. In contrast to the rapid inflation in the 
inflaton scalar field, our inflation is a power type, but
we have two independent curvature perturbations which make possible the consistency 
with the observed spectra.

For simplicity, on the other hand, I neglected the viscosity which may play important 
roles in dynamics 
between the outer space and the inner space. If we take the viscosity into account,
not only much entropy is produced (as shown in the previous paper\citep{tom}), but also
the severe condition such as $\lambda_2/\lambda_0 \simeq  (\lambda_1/\lambda_0)^2 
\ll 1$ for producing the observed CMB fluctuations in the scalar mode
may be softened. The next step is to study the perturbations and quantum 
fluctuations to derive the condition, in the case with viscous processes due to the 
transport of $10$-dimensional gravitational waves.\citep{tom, TI}

\appendix
\section{Higher-order terms $\Delta \Phi_h$ and $\Delta \Phi_H$ of curvature 
perturbations $\Phi_h$ and $\Phi_H$}
The higher-order terms $\Delta \Phi_h$ and $\Delta \Phi_H$ (with respect to 
$x$ and $y \ (\ll 1)$) in Eqs. (\ref{eq:c3}) 
and  (\ref{eq:c4})  are derived from the part of the Einstein equations in the scalar mode 
\begin{equation}
 \label{eq:aa1}
\delta R^0_i = - \delta T^0_i,
\end{equation} 
\begin{equation}
 \label{eq:aa2}
\delta R^0_a = - \delta T^0_a,
\end{equation} 
where the perturbed components of energy-momentum tensors are
\begin{equation}
 \label{eq:aa3}
\delta T^0_i = r(\rho+p)(v_s^{(0)}- b^{(0)}) q_i Q,
\end{equation} 
\begin{equation}
 \label{eq:aa4}
\delta T^0_a = R(\rho+p)(V_s^{(0)}- B^{(0)}) q Q_a.
\end{equation} 
Using the expressions of $\delta R^0_i$ and $\delta R^0_a$ in the Appendix of 
Abbott et al's paper\citep{abb}, we obtain
\begin{equation}
 \label{eq:aa5}
  \begin{split} 
-\frac{r}{k_r^{(0)}} (\rho+p) v_s^{(0)} &= -(d-1)\dot{\Phi}_h +[-dD\frac{\dot{R}}{R}-
(d-1)(d-2)\frac{\dot{r}}{r} ] \Phi_h \\
&- D\dot{\Phi}_H +[(2-d)D\frac{\dot{r}}{r}-D(D-1)\frac{\dot{R}}{R} ] \Phi_H \\
&- (\frac{k_R^{(0)}}{R})^2 \dot{\tilde{\Phi}}_G +[2(2-d)\frac{\dot{r}}{r}(\frac{k_R^{(0)}}{R})^2 -
D\frac{\dot{R}}{R}(\frac{k_r^{(0)}}{r})^2] \tilde{\Phi}_G \\
&+ \{\frac{1}{2}(\frac{k_R^{(0)}}{R})^2+(d-1) [\frac{\ddot{r}}{r} -(\frac{\dot{r}}{r})^2] 
+D \frac{\dot{r}}{r} \frac{\dot{R}}{R} \} (-\Phi_6 + \frac{r}{\dot{r}}\Phi_h
-\frac{R}{\dot{R}}\Phi_H),
 \end{split} 
\end{equation} 
\begin{equation}
 \label{eq:aa6}
  \begin{split} 
-\frac{R}{k_R^{(0)}} (\rho+p) V_s^{(0)} &= -(D-1)\dot{\Phi}_H +[-dD\frac{\dot{r}}{r}-
(D-1)(D-2)\frac{\dot{R}}{R} ] \Phi_H \\
&- d\dot{\Phi}_h +[(2-D)d\frac{\dot{R}}{R}-d(d-1)\frac{\dot{r}}{r} ] \Phi_h \\
&- (\frac{k_r^{(0)}}{r})^2 \dot{\tilde{\Phi}}_G +[2(2-D)\frac{\dot{R}}{R}(\frac{k_r^{(0)}}{r})^2 -
D\frac{\dot{r}}{r}(\frac{k_R^{(0)}}{R})^2] \tilde{\Phi}_G \\
&+ \{\frac{1}{2}(\frac{k_r^{(0)}}{r})^2+(D-1) [\frac{\ddot{R}}{R} -(\frac{\dot{R}}{R})^2] 
+d \frac{\dot{r}}{r} \frac{\dot{R}}{R} \} (\Phi_6 + \frac{R}{\dot{R}}\Phi_H
-\frac{r}{\dot{r}}\Phi_h),
 \end{split} 
\end{equation} 
where a dot denotes $d/dt$, and $\Phi_h, \Phi_H, \tilde{\Phi}_G$ and $\Phi_6$ are 
defined in Eqs.(15), (33) and (36) of [5]. At the final stage of the inflating outer space and 
the collapsing inner space, we have $\rho = p = 0$, as shown in Eq. (10) of [5]. So, the
left-hand sides of Eqs. (\ref{eq:aa5}) and  (\ref{eq:aa6}) vanish.

Now let us consider the stage of $x \ll 1$ and $y \ll 1$, where $x$ and $y$ are defined 
in Eqs.  (\ref{eq:a18}) and  (\ref{eq:a19}).  
Then, the lowest-order terms in $\Phi_h$ and $\Phi_H$ with respect to $x$ and $y$ 
are expressed as
\begin{equation}
 \label{eq:aa7}
\Phi_h = (\tau/\tau_i)^{-8/3} \Phi_{hi} , \quad \Phi_H = (\tau/\tau_i)^{-4/3} \Phi_{Hi},
\end{equation} 
and $\Phi_6 = \tilde{\Phi}_G = 0$.
To derive the next-order terms, let us put
\begin{equation}
 \label{eq:aa8}
\Phi_h = (\tau/\tau_i)^{-8/3} \Phi_{hi} + \Delta \Phi_h,
\end{equation} 
\begin{equation}
 \label{eq:aa9}
\Phi_H = (\tau/\tau_i)^{-4/3} \Phi_{Hi} + \Delta \Phi_H.
\end{equation} 
 From Eq.(32) of [5], we can derive
\begin{equation}
 \label{eq:aa10}
\tilde{\Phi}_G = \frac{3}{2} [-\tau^{-2/3} \tau_i^{8/3} \Phi_{hi} + \tau^{2/3} \tau_i^{4/3} 
\Phi_{Hi}] \ln (\tau/\tau_i),
\end{equation} 
\begin{equation}
 \label{eq:aa11}
(\tilde{\Phi}_G)' = \frac{3}{2} [-\tau^{-5/3} \tau_i^{8/3} \Phi_{hi} + \tau^{-1/3} \tau_i^{4/3} 
\Phi_{Hi}] + [\tau^{-5/3} \tau_i^{8/3} \Phi_{hi} + \tau^{-1/3} \tau_i^{4/3} 
\Phi_{Hi}] \ln (\tau/\tau_i),
\end{equation} 
where $\tau \equiv t_0 - t$ and a dash denotes $d/d\tau$. For $\Phi_6$, we have
\begin{equation}
 \label{eq:aa12}
\Phi_6' + \frac{1}{\tau}\Phi_6 = 3\tau (\Delta \Phi_h' + \frac{8/3}{\tau} \Delta \Phi_h
+\Delta \Phi_H' + \frac{4/3}{\tau} \Delta \Phi_H) +2[(k_r^{(0)}/r)^2 -(k_R^{(0)}/R)^2]
 \tilde{\Phi}_G,
\end{equation} 
which is derived from Eq.(58) of [5].
Here we define the auxiliary quantities $X$ and $Y$ by
\begin{equation}
 \label{eq:aa13}
X \equiv \Delta \Phi_h' + \frac{8}{3\tau} \Delta \Phi_h  \quad {\rm and} \quad 
Y \equiv \Delta \Phi_H' + \frac{4}{3\tau} \Delta \Phi_H.
\end{equation} 
Then, from Eqs. (\ref{eq:aa5}), (\ref{eq:aa8}), (\ref{eq:aa9}), (\ref{eq:aa11}) and (\ref{eq:aa12}), 
we obtain the following equations for $X$ and $Y$
\begin{equation}
 \label{eq:aa14}
2X + 6Y = A, 
\end{equation} 
and 
\begin{equation}
 \label{eq:aa15}
3X + 5Y = B + \frac{2}{\tau^2} \Phi_6,
\end{equation} 
where $A$ and $B$ are expressed as
\begin{equation}
 \label{eq:aa16}
   \begin{split} 
A &=  -3 (k_R^{(0)}/R_0)^2 \tau^{-1}  {\tau_i}^{4/3} \Phi_{Hi} + 
\{[-2 (k_R^{(0)}/R_0)^2 \tau^{-7/3} + 3(k_r^{(0)}/r_0)^2 \tau^{-1}] {\tau_i}^{8/3} \Phi_{hi} \\
 &-3 (k_r^{(0)}/r_0)^2 \tau^{1/3} {\tau_i}^{4/3} \Phi_{Hi} \} \ln (\tau/\tau_i),
   \end{split} 
\end{equation} 
and
\begin{equation}
 \label{eq:aa17}
 \begin{split} 
B &= 3(k_r^{(0)}/r_0)^2 \tau^{-1} {\tau_i}^{8/3} \Phi_{hi}
+\{[3(k_r^{(0)}/r_0)^2 \tau^{-1} - \frac{3}{2} (k_R^{(0)}/R_0)^2\tau^{-7/3}] {\tau_i}^{8/3} 
\Phi_{hi} \\
&+ [-5 (k_r^{(0)}/r_0)^2 \tau^{1/3} + \frac{3}{2} (k_R^{(0)}/R_0)^2\tau^{-1}] {\tau_i}^{4/3} 
\Phi_{Hi} \} \ln (\tau/\tau_i).
\end{split} 
\end{equation} 
Eliminating $\Phi_6$ from Eq.(\ref{eq:aa15}) by use of (\ref{eq:aa12}), we have
\begin{equation}
 \label{eq:aa18}
3X' + 5Y' + \frac{3}{\tau} (X + 3Y) = B' +\frac{3}{\tau} B + \frac{2}{\tau^2} C,
\end{equation} 
where $C$ is
\begin{equation}
 \label{eq:aa19}
C \equiv 2[(k_r^{(0)}/r)^2 - (k_R^{(0)}/R)^2] \tilde{\Phi}_G
\end{equation} 
with $\tilde{\Phi}_G$ defined by Eq.(\ref{eq:aa10}).

Integrating Eqs. (\ref{eq:aa14}) and (\ref{eq:aa18}) with respect to $X$ and $Y$, 
we obtain $X$ and $Y$, expressed as
\begin{equation}
 \label{eq:aa19a}
 \begin{split} 
X & = \tau^{-1} \{[\frac{3}{2}(\frac{k_r^{(0)}}{r_0})^2 -\frac{37}{28} \tau^{-4/3} 
(\frac{k_R^{(0)}}{R_0})^2] \ln (\tau/\tau_i) + [\frac{23}{24} (\frac{k_r^{(0)}}{r_0})^2 +
\frac{243}{392} \tau^{-4/3} (\frac{k_R^{(0)}}{R_0})^2] \} {\tau_i}^{8/3} \Phi_{hi} \\
&+ \tau^{1/3} \{[-12(\frac{k_r^{(0)}}{r_0})^2 +\frac{9}{4} \tau^{-4/3} 
(\frac{k_R^{(0)}}{R_0})^2] \ln (\tau/\tau_i) + [\frac{243}{8} (\frac{k_r^{(0)}}{r_0})^2 -
\frac{9}{4} \tau^{-4/3} (\frac{k_R^{(0)}}{R_0})^2] \} {\tau_i}^{4/3} \Phi_{Hi},
\end{split} 
\end{equation} 
and
\begin{equation}
 \label{eq:aa19b}
 \begin{split} 
Y & = \tau^{-1} \{\frac{3}{28} \tau^{-4/3} 
(\frac{k_R^{(0)}}{R_0})^2 \ln (\tau/\tau_i) - [\frac{23}{72} (\frac{k_r^{(0)}}{r_0})^2 +
\frac{81}{392} \tau^{-4/3} (\frac{k_R^{(0)}}{R_0})^2] \} {\tau_i}^{8/3} \Phi_{hi} \\
&+ \tau^{1/3} \{[\frac{7}{2}(\frac{k_r^{(0)}}{r_0})^2 -\frac{3}{4} \tau^{-4/3} 
(\frac{k_R^{(0)}}{R_0})^2] \ln (\tau/\tau_i) + [-\frac{81}{8} (\frac{k_r^{(0)}}{r_0})^2 +
\frac{1}{4} \tau^{-4/3} (\frac{k_R^{(0)}}{R_0})^2] \} {\tau_i}^{4/3} \Phi_{Hi}.
\end{split} 
\end{equation} 

Integrating Eq.(\ref{eq:aa13}) with respect to $\Delta \Phi_h$ and $\Delta \Phi_H$,
moreover, we obtain their following expressions 
\begin{equation}
 \label{eq:aa20}
 \begin{split} 
\Delta \Phi_h & = \{[\ln (\tau/\tau_i) +\frac{19}{72}] x^2 + [-\frac{37}{28} \ln (\tau/\tau_i) 
+\frac{939}{14\times 56}] y^2\} (\tau/\tau_i)^{-8/3} \Phi_{hi} \\
&+ \{-\frac{16}{3} [\ln (\tau/\tau_i) +\frac{89}{6}] x^2 + [\frac{3}{8} \ln (\tau/\tau_i) 
-\frac{1089}{64\times 49}] y^2\} (\tau/\tau_i)^{-4/3} \Phi_{Hi},
\end{split} 
\end{equation} 
and
\begin{equation}
 \label{eq:aa21}
 \begin{split} 
\Delta \Phi_H & = \{-\frac{23}{54} x^2 + [\frac{1}{42} (\ln \tau/\tau_i)^2  
-\frac{9}{112}] y^2 \} (\tau/\tau_i)^{-8/3} \Phi_{hi} \\
&+ \{[\frac{7}{3} \ln \tau/\tau_i -\frac{61}{8}] x^2 - [\frac{1}{12} \ln (\tau/\tau_i) 
+\frac{1}{48}] y^2 \} (\tau/\tau_i)^{-4/3} \Phi_{Hi} .
\end{split} 
\end{equation} 
For $x (\ll 1)$ and $y (\ll 1)$, therefore, $\Delta \Phi_h$ and $\Delta \Phi_H$ are
small, compared with the main terms $(\tau/\tau_i)^{-8/3} \Phi_{hi}$ and 
$(\tau/\tau_i)^{-4/3} \Phi_{Hi}$.


\end{document}